\documentclass[prd, aps, reprint, amsfonts, amssymb, amsmath, preprintnumbers, showpacs, nofootinbib, superscriptaddress]{revtex4-2}
\usepackage{graphicx, multirow,soul}
\usepackage[citecolor=blue]{hyperref}
\usepackage[all]{hypcap}
\usepackage{url,ulem}
\usepackage{slashed}
\usepackage{lipsum}
\usepackage{tikz}

\begin{document}

\title{Collapsing domain walls beyond $Z_2$}

\author{Yongcheng Wu}
\email{ycwu0830@gmail.com}
\affiliation{Department of Physics and Institute of Theoretical Physics, Nanjing Normal University, Nanjing, 210023, China}
\affiliation{Department of Physics, Oklahoma State University, Stillwater, OK 74078, USA}
\author{Ke-Pan Xie}
\email{kepan.xie@unl.edu}
\affiliation{Department of Physics and Astronomy, University of Nebraska, Lincoln, NE 68588, USA}
\author{Ye-Ling Zhou}
\email{zhouyeling@ucas.ac.cn (corresponding author)}
\affiliation{School of Fundamental Physics and Mathematical Sciences, Hangzhou Institute for Advanced Study, UCAS, Hangzhou, China}
\affiliation{International Centre for Theoretical Physics Asia-Pacific, Beijing/Hangzhou, China}

\date{\today}

\begin{abstract}
Discrete symmetries are widely imposed in particle theories. It is well-known that the spontaneous breaking of discrete symmetries leads to domain walls. Current studies of domain walls have focused on those from the spontaneous breaking of a $Z_2$ symmetry. Larger discrete symmetries have multiple degenerate vacua, leading to the domain walls in principle different from the simplest $Z_2$ domain wall. We take domain walls from $Z_N$ symmetry breaking as an illustrative study, and study in detail the $Z_3$ case, in which semi-analytical results for the tension and thickness of domain walls are derived. Explicit symmetry breaking terms lead to the dynamics of domain walls collapsing more complicated than the $Z_2$ case. Gravitational wave signals deviate from those from $Z_2$ domain walls.

\end{abstract}
\preprint{}
\maketitle

%%%%%%%%%%%%%%%%%%%%%%%%%%%%%%%%%%%%%%%%%%%%%%%%%%%%%%%%%%%%%%%%%%%%%%%%%%%%%%%
%%%%%%                          Introduction                             %%%%%%
%%%%%%%%%%%%%%%%%%%%%%%%%%%%%%%%%%%%%%%%%%%%%%%%%%%%%%%%%%%
\section{Introduction}

Discrete symmetries play important roles in particle physics. One of the most famous examples is the CP symmetry, which is an approximate $Z_2$ symmetry in the Standard Model (SM). In new physics, discrete symmetries are widely predicted or imposed to forbid unnecessary operators at very high energy scales. The D-parity symmetry appears in left-right symmetric and $SO(10)$ grand unified models \cite{Kibble:1982dd,Chang:1983fu}, the R-parity is introduced in supersymmetric models to suppress proton decay \cite{Ibanez:1991pr}, the $Z_N$ symmetry is applied in axion models to solve the strong CP problem \cite{Sikivie:1982qv}, and the either Abelian or non-Abelian discrete symmetries are imposed in the flavour space to address the flavour mixing patterns of quarks and leptons (see e.g. Refs.~\cite{King:2017guk,Xing:2019vks} for recent reviews).

In the early Universe, spontaneous breaking of a discrete symmetry generates degenerate vacua. These vacua are disconnected in the three dimensional space, and domain walls, which are two dimensional macroscopic objects in the core, form between them \cite{Kibble:1976sj}. The formation of domain walls is in general regarded as a problem in cosmology: once they form after inflation, they may soon dominate the energy density and overclose the Universe during the Hubble expansion \cite{Zeldovich:1974uw}. One way to solve this problem is to introduce bias terms in the potential \cite{Larsson:1996sp,Gelmini:1988sf,Vilenkin:1981zs} (see, e.g., \cite{Stojkovic:2005zh, King:2018fke} for other possible ways). These terms break the symmetry explicitly and make the vacua non-degenerate. This effect soon or later becomes significant as the Universe is cooled down to a temperature sufficiently lower than the scale of spontaneous symmetry breaking (SSB). Then domain walls collapse.
Following the wall collapsing is the production of gravitational waves (GWs) radiation, which form a stochastic background today.

Most studies on domain wall evolution and GW productions have assumed $Z_2$ as an illustrative example (see Ref.~\cite{Saikawa:2017hiv} for a recent review).
Lattice simulations are performed in Refs.~\cite{Hiramatsu:2010yz,Hiramatsu:2013qaa}, those including string-wall networks in $Z_N$-invariant axion models are performed in Ref.~\cite{Hiramatsu:2012sc,Kawasaki:2014sqa}, and that for a generic set of potentials is recently discussed in \cite{Krajewski:2021jje}. 
On the phenomenological side, the testability of spontaneous R-parity symmetry breaking using GWs is discussed in \cite{Dine:2010eb}.
Discussions on spontaneous CP violation at electroweak scale are given in \cite{Chen:2020wvu,Chen:2020soj}.
Domain walls following the first-order electroweak phase transition in the context of $Z_3$-invariant singlet-scalar-extended SM are discussed in~\cite{Zhou:2020ojf}.
What is more important is that GW detectors could therefore provide a characteristic signature for a large range of spontaneous discrete symmetry breaking scales that are hard to be tested in other experiments.
It was pointed out in \cite{Gelmini:2020bqg} that domain walls induced GW signal provides a potential way to test the origin of lepton flavour mixing. Depending on an adequately-chosen bias parameter, a large range of discrete flavour symmetry scale, from 1~TeV to $10^{14}$~GeV (the classical seesaw scale), can be potentially touched by the next-generation GW interferometers.
Axion-like particles as a dark matter candidate can be detected with mass from $10^{-16}$ to $10^6$~eV if they are produced at temperatures below 100 eV \cite{Gelmini:2021yzu}. Recent discussions on NANOGrav signal \cite{Arzoumanian:2020vkk} from collapsing domain walls in $Z_2$ and closed domain walls in the axion-like-particle models are respectively discussed in \cite{Bian:2020urb, Sakharov:2021dim}.

The purpose of this paper is to explore domain wall properties and the consequent GW signatures from discrete symmetries beyond $Z_2$. In section~\ref{sec:II}, we briefly review basic properties of the $Z_2$ domain walls. Section~\ref{sec:III} discusses properties of domain walls beyond $Z_2$. We take those from $Z_3$ symmetry for illustration in numerical simulation. The testability of discrete symmetries via GW detection is discussed in section~\ref{sec:IV}. We conclude in section~\ref{sec:V}.

\section{The simplest domain walls \label{sec:II}}
In the very early Universe, cosmic domain walls form during the SSB of discrete symmetries.
Most discussions on domain walls are based on a toy model of a real scalar in a global $Z_2$ symmetry. We give a brief review here following \cite{Vilenkin:1984ib}.

A tree-level $Z_2$-invariant potential of a real scalar $\phi$, up to an irrelevant constant term, is given by
\begin{eqnarray} \label{eq:potential_Z_2}
V_{Z_2} = -\frac{\mu^2}{2} \phi^2 + \frac{\lambda}{4} \phi^4\,,
\end{eqnarray}
where $\mu$, $\lambda>0$ are assumed without loss of generality. The potential is invariant under the transformation $\phi \to -\phi$.
The tree-level potential has two degenerate minima at $\phi = \pm v $  with $v = \mu/\sqrt{\lambda}$.

In the radiation-dominated era, the scalar interacts with the plasma, and the effective potential receives thermal corrections. In the high temperature limit, the potential is dominated by the thermal corrections, and the system stays in the symmetric vacuum $\langle \phi \rangle=0$. As temperature decreases during the Hubble expansion, the thermal term becomes less and less relevant. At some point, the tree-level terms dominate and the phase transition happens. As there is no preference among the degenerate vacua $\pm v$, $\phi$ may gain different vacuum expectation values (VEVs) in different spatial regions.
% Bubbles with different VEVs expand in the Universe.
% When they eventually meet with each other,
Then the energy barriers on the border form cosmic domain walls.

We call domain walls from $Z_2$ in Eq.~\eqref{eq:potential_Z_2} the {\it simplest domain walls}.
By fixing the VEV $\langle \phi \rangle =\pm v$ at $z=\pm\infty$, where $z$ is the coordinate of axis perpendicular to the wall and $z=0$ as the centre of the wall,
the solution of simplest domain walls is obtained by solving the equation of motion (EOM) $\phi''(z) = \partial V(\phi) / \partial \phi$, where ``$\,\prime\,$'' denotes the derivative with respect to $z$,
\begin{eqnarray}\label{Z2_solution}
\phi = v \tanh\left( \sqrt{\frac{\lambda}{2}} v z \right) \,.
\end{eqnarray}
Here $\delta \approx \sqrt{2/(\lambda v^2)}$ is regarded as the thickness of the wall, which estimates the typical length scale of the scale variation of $\phi(z)$.

The energy stored per unit area on the wall, also called the tension of the wall, is calculated via the energy momentum tensor, $T_{\mu\nu} =  \partial_\mu \phi \partial_\nu \phi - {\cal L} g_{\mu\nu}$. Along the direction perpendicular to the wall, the $(0,0)$ entry, i.e. the energy density component, is
\begin{eqnarray}
\varepsilon (z) \equiv T_{00}= \frac{1}{2} \Big[\phi'(z)\Big]^2 + \Delta V(\phi(z))\,,
\end{eqnarray}
where $\Delta V(\phi) = V(\phi) - V_{\rm min}$.
The integration along $z$ gives the tension of the wall $\sigma = \int_{-\infty}^{\infty} d z \varepsilon(z)$. For the simplest domain wall, it is calculated to be $\sigma = \frac{4}{3} \sqrt{\frac{\lambda}{2}} v^3$.

%%%%%%%%%%%%%%%%%

\section{Beyond the simplest domain walls \label{sec:III}}

\begin{figure*}
\centering
\includegraphics[width=.75\textwidth]{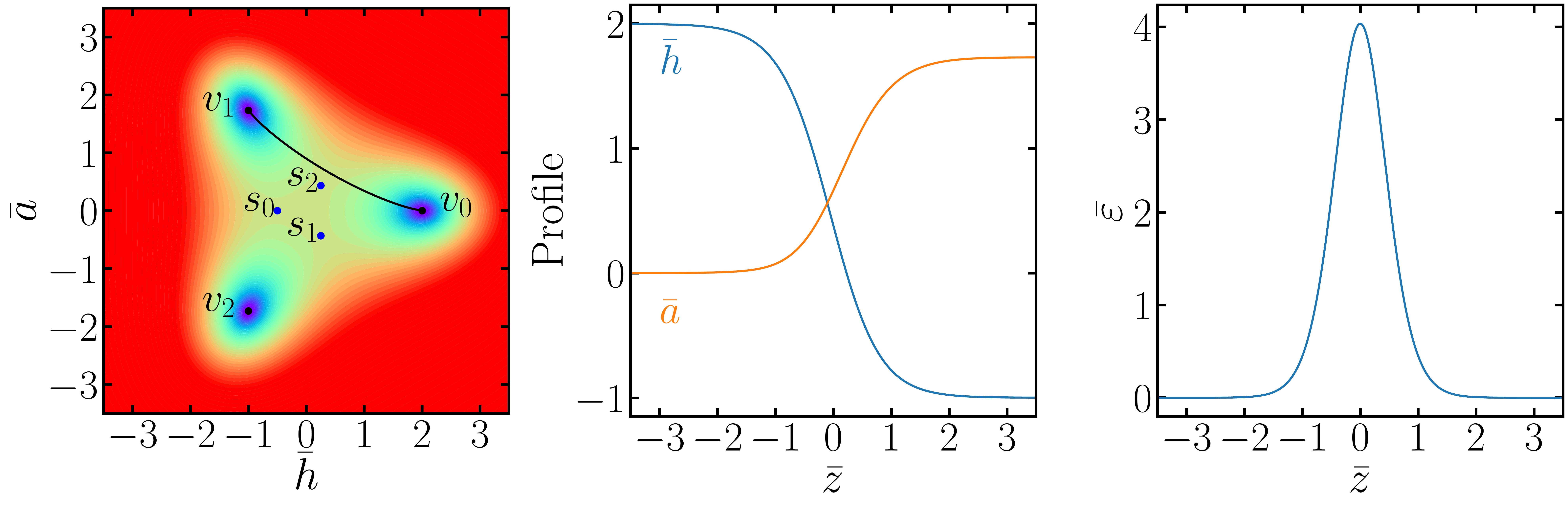}
\caption{Contour plot of the Coca-Cola-bottle-like potential (left panel), domain wall solution (middle panel) and energy density in the wall (right panel) in renormalisable $Z_3$ symmetry breaking. Vacua and saddle points are shown in the left panel. The path from $v_0$ to $v_1$ refers to the domain wall solution shown in the middle panel for $z$ from $-\infty$ and $+\infty$. $\mu/\sqrt{\lambda_1}$ is normalised to one, and $\beta =3/4$ is used.}\label{fig:DW_Z3}
\end{figure*}

We extend the discussion from $Z_2$ to $Z_N$ with $N \geqslant 2$. Furthermore, as CP symmetry is only slightly broken in the SM, it is reasonable to impose it here as a leading-order approximation. Explicit-breaking terms will not be considered in this section.

{\bf $Z_N$-invariant theories}. The first step is to extend the real scalar to a complex scalar $\phi=(h+ ia)/\sqrt{2}$. Perform a $Z_N$ transformation
\begin{eqnarray}
T: \,\phi \to e^{i 2\pi/N} \phi\,.
\end{eqnarray}
The $Z_N$ symmetry allows operators of $\phi$ to be $\phi^* \phi$, $\phi^N$, $\phi^{* N}$ or their products. Under the CP transformation,
\begin{eqnarray}
S: \,\phi \to \phi^*\,,
\end{eqnarray}
$\phi^N$ and $\phi^{* N}$ are enforced to appear as combinations of $\phi^N+\phi^{* N}$. The transformations $T$ and $S$ satisfy $T^N = S^2 = (TS)^2 =1$. While $T$ results in a rotation of $2\pi/N$ on the complex plane of the field $\phi$, $S$ is a reflection between the positive and negative imaginary parts transformation. They generate the dihedral group $D_N \simeq Z_N \rtimes Z_2$ and also denoted as $\Delta(2N)$ (see e.g., Ref. \cite{Ishimori:2010au} for a review of discrete symmetries). Here, the parity symmetry $Z_2$ represents the CP symmetry imposed in the theory.   We emphasise that $D_N$ is a natural consequence of the Abelian discrete symmetry $Z_N$ and the CP symmetry.

In general, the $Z_N$- and CP-invariant potential for a complex scalar $\phi$ must take the form
\begin{eqnarray} \label{eq:potential_Z_N}
{V}_{Z_N} = f\left(\phi^* \phi, \phi^N +\phi^{* N}\right) \,.
\end{eqnarray}
The most simplified version of the potential might be
\begin{eqnarray}
{V}_{Z_N} &=&  -\mu^2 \phi^* \phi + \lambda_1 (\phi^* \phi)^2 - \lambda_2 \mu^{4-N} (\phi^N+\phi^{*N})\,, \label{eq:potential_Z_N}
\end{eqnarray}
where all coefficients are real, $\mu$, $\lambda_1$, $\lambda_2 >0$ are assumed without loss of generality.\footnote{In fact, the CP symmetry in this kind of potential does not have to be imposed but can be accidental. Given the more general one,
$-\mu^2 \phi^* \phi + \lambda_1 (\phi^* \phi)^2 - \lambda_2 \mu^{4-N} (e^{i\alpha}\phi^N+e^{-i\alpha}\phi^{*N})$.
With the phase rotation $\phi \to e^{i\alpha/N} \phi$, we arrive at Eq.~\eqref{eq:potential_Z_N}. With the phase rotation, one can further keep $\lambda_2>0$. Note that the phase rotation may induce a CP violation in the coupling for $\phi$ with other particles, which does not effect the scalar potential property at leading order.} In this potential, the $\lambda_2$ term is the only source to break the $U(1)$ symmetry. For $N=3$ or 4, the potential is the most-general $Z_N$-invariant renormalisable potential at tree level. For $N \geqslant 5$, it becomes non-renormalisable, which can be realised by introducing additional heavy particles. The global $U(1)$ symmetry is restored in the limit $\lambda_2 \to 0$.
Back to the general potential in Eq.~\eqref{eq:potential_Z_N}, it takes a general feature in the field space that the local maximal point is located in the central and surrounded by $N$ minima, just like the ``bottom of a classical Coca-Cola bottle''. It can be easily seen by parametrising $\phi = \rho\, e^{i\theta}$ and ${V}_{Z_N}$ is then written in the form ${V}_{Z_N} = f(\rho, \cos (N\theta))$. The minima, referring to $N$ degenerate vacua of $Z_N$, are given by $\langle \phi \rangle = v_0e^{i2\pi k /N}$ where $k=0$, 1, ..., $N-1$ and $v_0$ is the solution of $\partial_\rho f(\rho, \cos (N\theta))|_{\rho=v_0,\, \theta=0} = 0$.

In general, domain walls refer to solutions of the scalar profile across different vacua, i.e., from $v_0e^{i2\pi k_1/N}$ to $v_0e^{i2\pi k_2/N}$ with $k_1 \neq k_2$. A general discussion on domain walls of $Z_N$ is complicated. We will take $Z_3$ as an illustrative case to explore general futures of domain walls that are not shared by $Z_2$ domain walls. Explicit-breaking terms (i.e. bias terms) will be considered in next section.

{\bf The $Z_3$ case}. The renormalisable $Z_3$-invariant potential of $\phi$ is generically given by
\begin{eqnarray} \label{eq:potential_Z_3}
V_{Z_3} &=&  -\mu^2 \phi^* \phi + \lambda_1 (\phi^* \phi)^2 - \lambda_2 \mu (\phi^3+\phi^{*3})\,,
\end{eqnarray}
where all coefficients are real and positive. Here, the CP symmetry does not have to be imposed but accidentally  conserved due to the requirement of renormalisation and Hermiticity of the potential.
Thus, the symmetry is enlarged to $D_3$.
%By minimising the potential in Eq.~\eqref{eq:potential_Z_3},

There are three degenerate vacua, explicitly written as
\begin{eqnarray} \label{eq:vev}
\langle \phi \rangle \equiv v_k = \frac{\mu}{\sqrt{2\lambda_1}} (\beta + \sqrt{1+\beta^2}) \, e^{i2\pi k /3}
\end{eqnarray}
for $k=0,1,2$, where $\beta = 3\lambda_2/\sqrt{8\lambda_1}>0$. In addition, there are three saddle points,
\begin{eqnarray}
s_k = \frac{\mu}{\sqrt{2\lambda_1}} (\beta - \sqrt{1+ \beta^2}) \, e^{i2\pi k /3}
\end{eqnarray}
These solutions are shown in the left panel in Fig.~\ref{fig:DW_Z3} for $\beta = 3/4$. %Note that in the alternative case where $\lambda_2<0$ (i.e., $\beta<0$), $s_k$ appear as vacua and $v_k$ appear as saddle points in the potential.
The two real components of $\phi$ gain masses after the SSB. The model parameters match with the physical observables via
\begin{eqnarray}
\mu^2 &=& \frac{1}{6} (3m_h^2 - m_a^2) \,, \nonumber\\
\lambda_1 &=& \frac{1}{12 v_0^2} (3m_h^2 + m_a^2) \,, \\
\beta &=& \frac{m_a^2}{\sqrt{9m_h^4 - m_a^4}} \,,\nonumber
\end{eqnarray}
where $v_0$ refers to Eq.~\eqref{eq:vev} at $k=0$.
Here $m_h$ and $m_a$ are the masses of $h$ and $a$ respectively generated after SSB.
% Here $m_h$ and $m_a$ are mass eigenvalues generated after SSB.
% They are simply masses of $h$ and $a$ in the vacuum $v_0$, but differed by a phase rotation in the other vacua.

\begin{figure*}
\centering
\includegraphics[width=.65\textwidth]{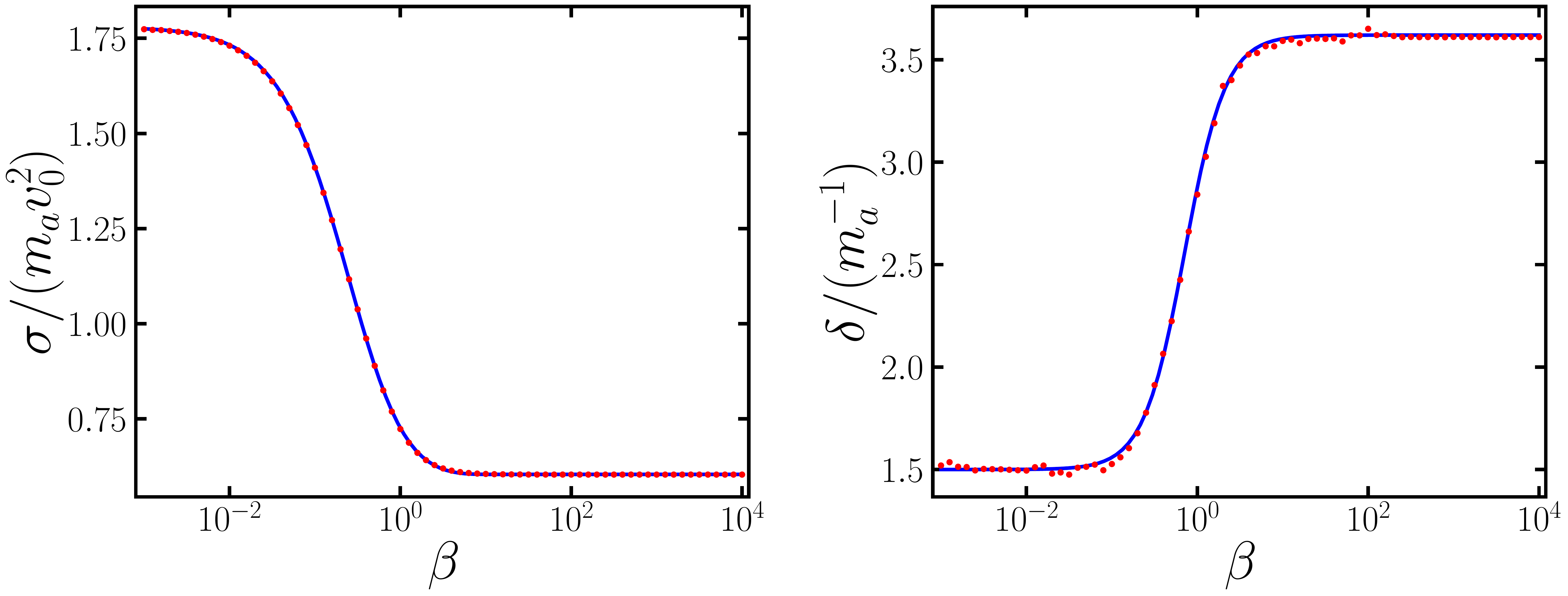}
\caption{Dependence of the domain wall tension $\sigma$ and thickness $\delta$ on $\beta$.}\label{fig:Z3_sigma_width}
\end{figure*}

{\bf $Z_3$ domain walls}. By fixing $v_1$ at $z=-\infty$ and $v_0$ at $z= +\infty$, we are able to obtain the domain wall solution by solving the EOM
$\phi_i''(z) = \partial V(\phi) / \partial \phi_i$
for $\phi_i = h$, $a$. This problem can be solved be performing a normalisation of the field and coordinate $\bar{h} = \frac{\sqrt{\lambda_1}}{\mu} h$, $\bar{a} = \frac{\sqrt{\lambda_1}}{\mu} a$ and $\bar{z} = \mu \, z$. In this case all variables are dimensionless and the system depends only on one parameter $\beta$.

Given numerical values for the free parameters in the potential, the EOM can be solved by adapting a modified version \cite{Chen:2020wvu,Chen:2020soj} of the path deformation algorithm \cite{Wainwright:2011kj}, where the true path between the two vacua is obtained iteratively by deforming the path according to the ``force'' felt along the path. Along the path, where the system is effectively reduced to a one-dimension problem, a 5-th order Runge-Kutta method is used to solve the corresponding EOM. We find that, different from the scalar profile $\phi_i(z)$ derived for the $Z_2$ domain walls shown in Eq.~(\ref{Z2_solution})~\cite{Vilenkin:1984ib}, scalar profiles cannot be simply fitted by a hyperbolic tangent profile. In the middle panel of Fig.~\ref{fig:DW_Z3}, we show an example of normalised scalar profiles as a function of the normalised coordinate with $\beta = 3/4$.

The tension of the wall is derived to be $\sigma = \int_{-\infty}^{\infty} d z \varepsilon(z) \equiv \frac{\mu^3}{\lambda_1} \bar{\sigma}$, where
\begin{eqnarray}
\bar{\sigma} &=& \int_{-\infty}^{\infty} d \bar{z} \left[ \frac{1}{2} (\bar{h}^{\prime 2} + \bar{a}^{\prime 2}) + \Delta \bar{V} \right]
\end{eqnarray}
is the normalised tension respect to the normalised fields and coordinate and $\Delta \bar{V} =\Delta V \lambda_1/\mu^4 $.
As $\bar{\sigma}$ only depends on one free parameter $\beta$, we calculated $\bar{\sigma}$ by scanning $\beta$ in a wide range $10^{-3}\leqslant \beta \leqslant 10^4$.
It is convenient to represent the tension in terms of physical observables,
\begin{eqnarray}
\sigma = m_a v_0^2 f(\beta) \,,
\end{eqnarray}
where $f(\beta)$ is a dimensionless order-one factor. The dependence of $f(\beta)$ on $\beta$ is fitted by a semi-analytical functions
\begin{eqnarray}
f(\beta) &=& 0.604 + \frac{0.234}{e^{0.826 \beta} + 0.435 \beta^2 - 0.801} \,.
\end{eqnarray}
The numerical data points and fitting function are both shown in the left panel of Fig.~\ref{fig:Z3_sigma_width}.
We guarantee that this formula matches with the numerical results very well with a relative error less than $2\%$ for $10^{-3} \leqslant \beta \leqslant 10^4$.

One can also calculate the thickness of the wall. We generalise the definition of the thickness as follows,\footnote{In the special case of $Z_2$ domain wall \cite{Vilenkin:1984ib}, $\delta$ is defined as the factor appearing in the a hyperbolic tangent function of the scalar profile $ \propto \tanh(z/\delta)$. This definition leads to the condition in Eq.~\eqref{eq:thickness_def}. We apply this condition as a generalised definition of the wall thickness for walls without a hyperbolic profile.}
\begin{eqnarray} \label{eq:thickness_def}
\int_{-\delta/2}^{\delta/2} dz \varepsilon(z) = 64\% \times \sigma.
\end{eqnarray}
It is parameterised in terms of the inverse of the scalar mass $m_a$ and an $\beta$-dependent order-one function,
\begin{eqnarray}
\delta = m_a^{-1} g(\beta)\,,
\end{eqnarray}
where the $\beta$-dependent function $g(\beta)$ is semi-analytically fitted by
\begin{eqnarray}
g(\beta) &=& 3.62-\frac{2.12}{1+1.85 \beta^{1.81}}\,,
\end{eqnarray}
with the same precision as the fit of $f(\beta)$.
We distinguish two cases $\beta \ll 1$ and $\beta \gtrsim 1$ and make comments below.

As the $\beta$-dependent term is the only source to give rise to the soft breaking of $U(1)$ symmetry, a global $U(1)$ is restored in the limit $\beta \to 0$. The spontaneous breaking of the approximate $U(1)$ leads to a pseudo-Nambu-Goldstone boson with mass $m_a^2 \simeq 3 \beta m_h^2 \ll m_h^2$. In the case $\beta \ll 1$, i.e., $\lambda_2 \ll \sqrt{\lambda_1}$, we encounter a two-step SSB as the temperature decreases during the Hubble expansion.
The first step is the SSB of $U(1)$, happening around the scale $v_0 \simeq \mu/\sqrt{2\lambda_1}$. At this scale, the first and second terms of the potential in Eq.~\eqref{eq:potential_Z_3} dominate the symmetry breaking. Shortly after, $\phi$ gains a VEV with the absolute value $v_0$ and an arbitrary phase in $[0, 2 \pi)$.
A well-known consequence following the $U(1)$ breaking is the formation of cosmic strings. The string tension is given by $\pi v_0^2$ up to an order-one factor \cite{Hindmarsh:1994re}.
Later when the temperature decreases into the energy scale comparable with $\sqrt{m_a M_{\rm P}}$ with $M_{\rm P}$ the Planck mass, the third term of Eq.~\eqref{eq:potential_Z_3} becomes non-negligible with the gradient energy, which is of order $v_0^2 H^2$ and $H$ the Hubble parameter, and the SSB of $Z_3$ begins to happen. The phase of $\phi$'s VEV is then fixed to one of the three phases $0$, $2\pi/3$ and $4\pi/3$. The energy barrier between spatial regions with different phases soon forms a domain wall with tension $\sigma \simeq 2.18 \sqrt{\beta} \mu^3/\lambda_1 \simeq 1.8 m_a v_0^2$. Each wall on its boundary attaches to a string and a topological defect of walls bounded by strings forms \cite{Kibble:1982dd, Everett:1982nm}.

This picture has a lot of phenomenological applications. For example, it can be applied to neutrino mass models with $U(1)_{B-L}$ symmetry broken to $Z_3$ and flavour models with Froggatt-Nielson mechanism broken to discrete symmetries. Cosmological consequences of domain walls in these models will be considered elsewhere.
Another widely studied example is in axion models with a $U(1)$ Peccei-Quinn symmetry explicitly broken to $Z_N$ with $N>1$, where $N$ is domain wall number connecting to a string (see e.g., \cite{Vilenkin:1984ib} for a review). We have checked that our results of tensions and thickness match with those in axion models by taking $N=3$ \cite{Hiramatsu:2012sc}.

$\beta \gtrsim 1$ is a region that has not been studied before. In this region, $U(1)$ should not  be regarded as an approximate symmetry any more. When the temperature decreases to the energy scale comparable with $v_0$, $\phi$ gains VEVs in different spatial regions and $Z_3$ is directly broken. Soon domain walls form on the barriers between different vacua. The tension satisfies a good approximation $\sigma \simeq m_h v_0^2$ and in particular, $\sigma \to m_h v_0^2$ for $\beta \to \infty$ (Note that $m_a^2 = 3 m_h^2/\sqrt{1+\beta^{-2}}$). In order to keep the theory perturbatively safe from radiative correction, two scalar masses must satisfy the condition $m_h$, $m_a \lesssim v_0/\beta$.

\section{Gravitational waves from collapsing walls \label{sec:IV}}

Although the formation of domain walls is ubiquitous in new physics models, long-lived domain walls will dominate the Universe and cause accelerating expansion that is already ruled out by current observations. The most popular and natural way to solve this problem is to add explicit breaking terms to the potential. These terms induce energy biases between vacua and make them not strictly degenerate.

A typical example of $Z_3$ explicit breaking which does not shift vacua but generate energy bias among them can be written as
\begin{eqnarray}
V_{\slashed{Z}_3} =\frac{2e^{i\alpha}}{3\sqrt{3}} \epsilon \phi \left( \frac{1}{4} \phi^3 - v_0^3 \right) + {\rm h.c.} \,,
\end{eqnarray}
where $\alpha$ is a free parameter. The vacuum energy difference between $v_i$ and $v_j$ is defined as $(V_{\rm bias})_{ij} = V_{v_i} - V_{v_j}$, and we have
\begin{equation}\begin{split}
(V_{\rm bias})_{10}=&~\epsilon v_0^4\cos\left(\alpha+\frac\pi6\right),\\
 (V_{\rm bias})_{20}=&~\epsilon v_0^4\cos\left(\alpha-\frac\pi6\right).
\end{split}\end{equation}
Without loss of generality, we assume $\alpha\in(0,\pi/3)$ so that  $(V_{\rm bias})_{20}>(V_{\rm bias})_{10}>0$ and hence the potential gains the global minimum at $v_0$, while $v_1$ and $v_2$ are false vacua with $v_2$ being the highest energy state. In the case of $\alpha=0$, the CP is not broken in the potential, we obtain two degenerate biases $(V_{\rm bias})_{10} = (V_{\rm bias})_{20}$ and $(V_{\rm bias})_{21} = 0$.
In the generic case with a nonzero $\alpha$, we arrive at three different biases which satisfy $(V_{\rm bias})_{20} - (V_{\rm bias})_{10} = (V_{\rm bias})_{21}$.

Due to the existence of energy biases, the domain walls will collapse and annihilate at late time. The annihilation happens when the tension $p_T\sim\mathcal{A}\sigma/t$~\cite{Press:1989yh,Garagounis:2002kt,Avelino:2005kn} decreases to be comparable with the vacuum pressure caused by the $V_{\rm bias}$'s. In the case of $Z_2$ domain wall collapsing, there is only one type of walls separating the true vacuum and the false vacuum, and thus only a single $V_{\rm bias}$ exists. The temperature for wall annihilation is given by
\begin{multline} \label{eq:T_ann}
T_{\rm ann}=3.41\times10^{-2}~{\rm GeV}\times C_{\rm ann}^{-1/2}\mathcal{A}^{-1/2}\times\\
\left(\frac{10}{g_*(T_{\rm ann})}\right)^{1/4}
\left(\frac{\rm TeV}{\sigma^{1/3}}\right)^{3/2}\left(\frac{V_{\rm bias}^{1/4}}{\rm MeV}\right)^2,
\end{multline}
where $\mathcal{A}$ and $C_{\rm ann}$ are $\mathcal{O}(1)$ constants related to the $N$ of the $Z_N$ symmetry. We take $\mathcal{A}=1.10 \pm 0.20$ and $C_{\rm ann}=5.02 \pm 0.44$ based on the simulation of axion models with $N=3$ \cite{Kawasaki:2014sqa}. Requirement that domain walls annihilate before they dominate the Universe sets a lower bound on the bias
\begin{equation}
V_{\rm bias}>C_{\rm ann}\frac{32\pi\mathcal{A}^2\sigma^2}{3M_{\rm Pl}^2}.
\end{equation}
On the other hand, domain walls will only form when the bias is smaller than some value set by percolation theory \cite{Stauffer:1978kr}:
\begin{align}
V_{\rm bias} < V_b\times \log\frac{1-p_c}{p_c}\approx 0.795 V_b\,,
\end{align}
where $V_b = \frac{4}{9}\big[\sqrt{1+\beta^2}\big(\sqrt{1+\beta^2}-\beta\big)\big]^3m_a^2v_0^2$ is the barrier hight of the potential between two minima.
Lacking a detailed simulation for the $Z_3$ domain wall annihilation, we will use the $Z_2$ results given above to estimate the features of the $Z_3$ domain wall collapse.

\begin{figure*}
\centering
\includegraphics[width=0.6\textwidth]{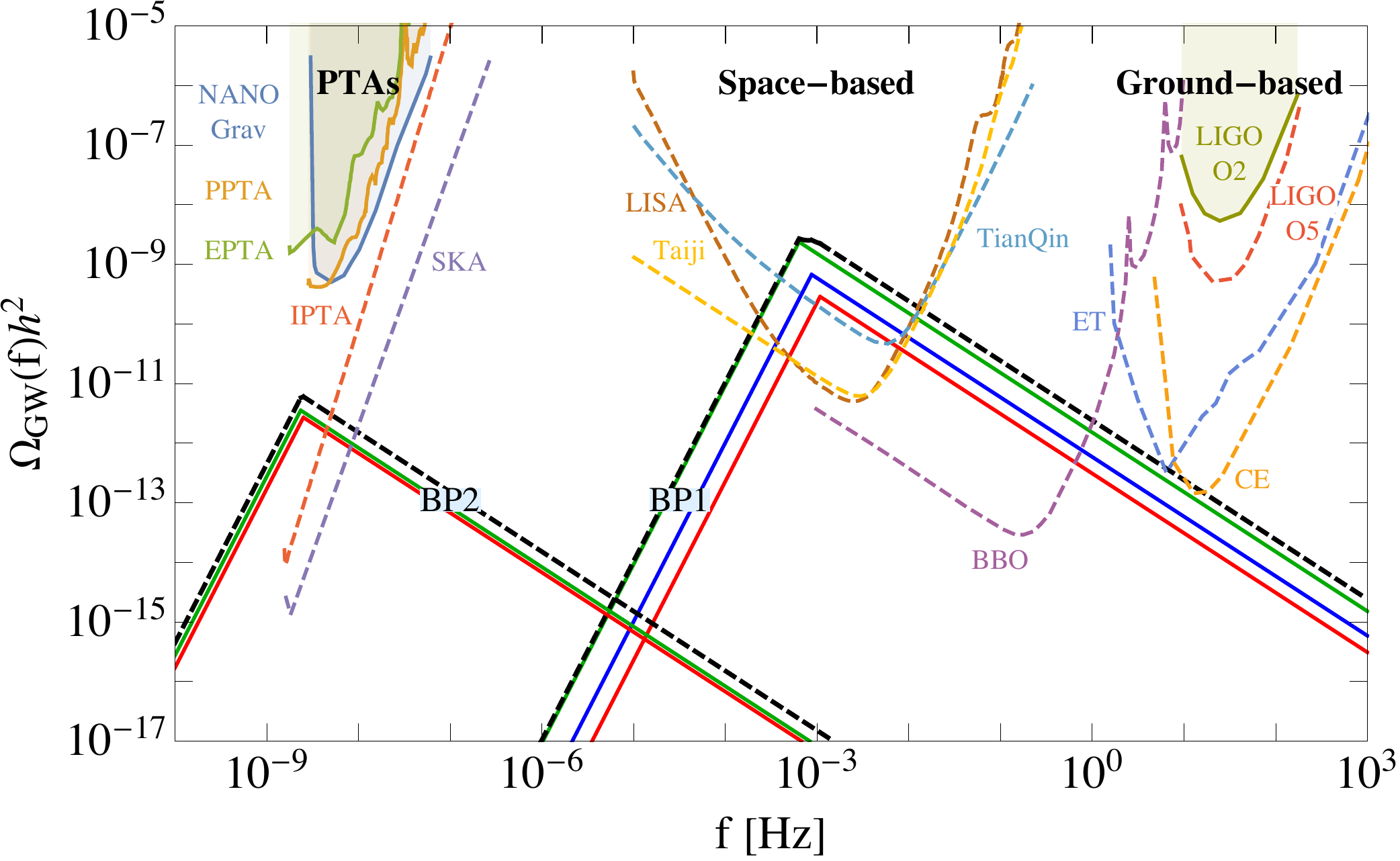}
\caption{Illustration for the GW spectrum from $Z_3$ domain wall collapse. For a given benchmark (BP), the red, green and blue curves represent the GW contribution from $(V_{\rm bias})_{20}$-, $(V_{\rm bias})_{10}$- and $(V_{\rm bias})_{21}$-driven annihilations, respectively, while the black dashed line is the total signal. The BPs are chosen by fixing $\beta=1$ and $\epsilon=10^{-28}$, and for BP1, $v_0=10^{11}$ GeV, $m_a=2$ TeV, $\alpha=2\pi/9$; while for BP2, $v_0=10^5$ GeV, $m_a=500$ GeV, $\alpha=\pi/27$. BP1 belongs to the [(01)2] scenario thus has three contributions to the GW spectrum, while BP2 belongs to the [0(12)] scenario whose GW spectrum has only two contributions.}\label{fig:gw_f}
\end{figure*}

In the case of $Z_3$ domain wall collapsing, we encounter three $V_{\rm bias}$'s, two of which are independent. They can be classified into two kinds. The first kind includes $(V_{\rm bias})_{10}$ and $(V_{\rm bias})_{20}$, which refer to walls separating false vacua and the true vacuum. The second kind is $(V_{\rm bias})_{21}$, corresponding to walls separating two different false vacua. Given the vacuum energy ordering $V_{v_0}<V_{v_1}<V_{v_2}$, there are two typical scenarios for the bias energy hierarchy:
\begin{enumerate}
\item The [(01)2] scenario. It represents the true vacuum is almost degenerate with the first false vacuum, and a large gap with the second false vacuum, $V_{v_0} \lesssim V_{v_1} \ll V_{v_2}$. Biases in this scenario follow the hierarchy $(V_{\rm bias})_{10}\ll (V_{\rm bias})_{21} \lesssim (V_{\rm bias})_{20}$;
\item The [0(12)] scenario. Two false vacua are nearly degenerate and their vacuum energy is much larger than the true vacuum $V_{v_0} \ll V_{v_1} \lesssim V_{v_2}$. Biases follow the hierarchy $(V_{\rm bias})_{10}\lesssim (V_{\rm bias})_{20} \gg  (V_{\rm bias})_{21}$.
\end{enumerate}
The two scenarios are sketched as follows.
\begin{equation*}
[(01)2]:\begin{array}{c}\rule[2pt]{7mm}{0.05em} V_{v_2} \rule[2pt]{7mm}{0.05em}\\
\\
\\
\rule[2pt]{7mm}{0.05em}V_{v_1}\rule[2pt]{7mm}{0.05em}\\
\rule[2pt]{7mm}{0.05em}V_{v_0}\rule[2pt]{7mm}{0.05em}\end{array}
;\qquad
[0(12)]:\begin{array}{c}
\rule[2pt]{7mm}{0.05em}V_{v_2}\rule[2pt]{7mm}{0.05em}\\
\rule[2pt]{7mm}{0.05em}V_{v_1}\rule[2pt]{7mm}{0.05em}\\
\\
\\
\rule[2pt]{7mm}{0.05em}V_{v_0}\rule[2pt]{7mm}{0.05em}
\end{array}
\end{equation*}
Following the behaviour $T_{\rm ann} \propto \sqrt{V_{\rm bias}}$, walls collapse at different temperatures for different biases. Those with the largest bias annihilate first. This leads to different evolution paths for the two different bias energy hierarchy scenarios.

In both scenarios, the annihilation driven by $(V_{\rm bias})_{20}$ happens first. After that, the remaining vacua are basically $V_{v_0}$ and $V_{v_1}$, but there could be $V_{v_2}$ remnants surrounding by $V_{v_1}$. The $V_{v_1}$ vacuum region could act as a screening medium which protects the $V_{v_2}$ vacuum from being ``eaten'' by $V_{v_0}$. As the temperature further decreases due to Hubble expansion, In the [(01)2] scenario, the next step is the annihilation driven by $(V_{\rm bias})_{21}$, and then the annihilation of walls between $(V_{\rm bias})_{10}$. Therefore, we get a three-step annihilation pattern. In the [0(12)] scenario, however, the situation changes, as the annihilation of $(V_{\rm bias})_{10}$ happens first, and the $V_{v_2}$ remnants included in the $V_{v_1}$ regions are annihilated at the same time. Therefore, in the [0(12)] scenario, there is no $(V_{\rm bias})_{21}$-driven annihilation, and an only two-step annihilation pattern is obtained.

The collapse and annihilation of domain walls produces GWs. We apply the results in $Z_2$ domain walls as an estimation for the $Z_3$ case, where the peak frequency as well as the energy density today are respectively~\cite{Saikawa:2017hiv}
\begin{multline}
f_{\rm peak}=1.1\times10^{-9}~{\rm Hz}\times\left(\frac{g_*(T_{\rm ann})}{10}\right)^{1/2}\times\\
\left(\frac{10}{g_{*S}(T_{\rm ann})}\right)^{1/3}\left(\frac{T_{\rm ann}}{10^{-2}~{\rm GeV}}\right),
\end{multline}
and
\begin{multline}
\Omega_{\rm GW}(f_{\rm peak})h^2=7.2\times10^{-18}\times\tilde\epsilon_{\rm GW}\mathcal{A}^2\times\\
\left(\frac{10}{g_*(T_{\rm ann})}\right)^{4/3}\left(\frac{\sigma^{1/3}}{\rm TeV}\right)^6\left(\frac{10^{-2}~{\rm GeV}}{T_{\rm ann}}\right)^4,
\end{multline}
where $\tilde\epsilon_{\rm GW}\approx0.7\pm0.4$~\cite{Hiramatsu:2013qaa}. While the full GW spectrum is difficult to obtain due to the limitation of simulation dynamical ranges, one can use $\Omega_{\rm GW}h^2\propto f^3$ for $f<f_{\rm peak}$ and $\Omega_{\rm GW}h^2\propto f^{-1}$ for $f>f_{\rm peak}$ to approximate the spectrum. From the discussion of $V_{\rm bias}$ we can infer that the GW spectrum of the [(01)2] and [0(12)] scenarios receives three and two separated contributions from domain wall annihilation, respectively.

Figure~\ref{fig:gw_f} shows two benchmark points of the GW spectrum from $Z_3$ domain wall annihilation, where $\beta=1$ and $\epsilon=10^{-28}$ are fixed, and we further adopt
\begin{equation}\begin{split}
{\rm BP1}:&~v_0=10^{11}~{\rm GeV},\quad m_a=2~{\rm TeV},\quad \alpha=\frac{2\pi}{9};\\
{\rm BP2}:&~v_0=10^{5}~{\rm GeV},\quad m_a=500~{\rm GeV},\quad \alpha=\frac{\pi}{27}.
\end{split}\end{equation}
The difference choices of the $\alpha$ parameter choice makes BP1 belong to the [(01)2] scenario and BP1 belong to the [0(12)] scenario. Therefore, the GW spectrum of BP1 receives three contributions from the three-stop annihilation pattern; while the spectrum of BP2 has only two contributions from the two-step annihilation pattern, as shown in the figure. We can see that the total shape of signal is always dominated by the component that has the lowest annihilation temperature. In general, for the GWs from $Z_N$ domain walls collapse, if the contribution from several $V_{\rm bias}$'s are comparable, we should find a multi-peak structure in the spectrum when zooming in the peak. For example, in Fig.~\ref{fig:gw_f} there is a mild two-peak shape for the signal spectrum. However, those multi-peak structure is in general not prominent and might be very challenging to detect in a real experiment. Therefore, the GWs from $Z_N$ domain wall collapse do not show a significant shape compared with the $Z_2$ case.

\begin{figure*}
\centering
\includegraphics[width=\textwidth]{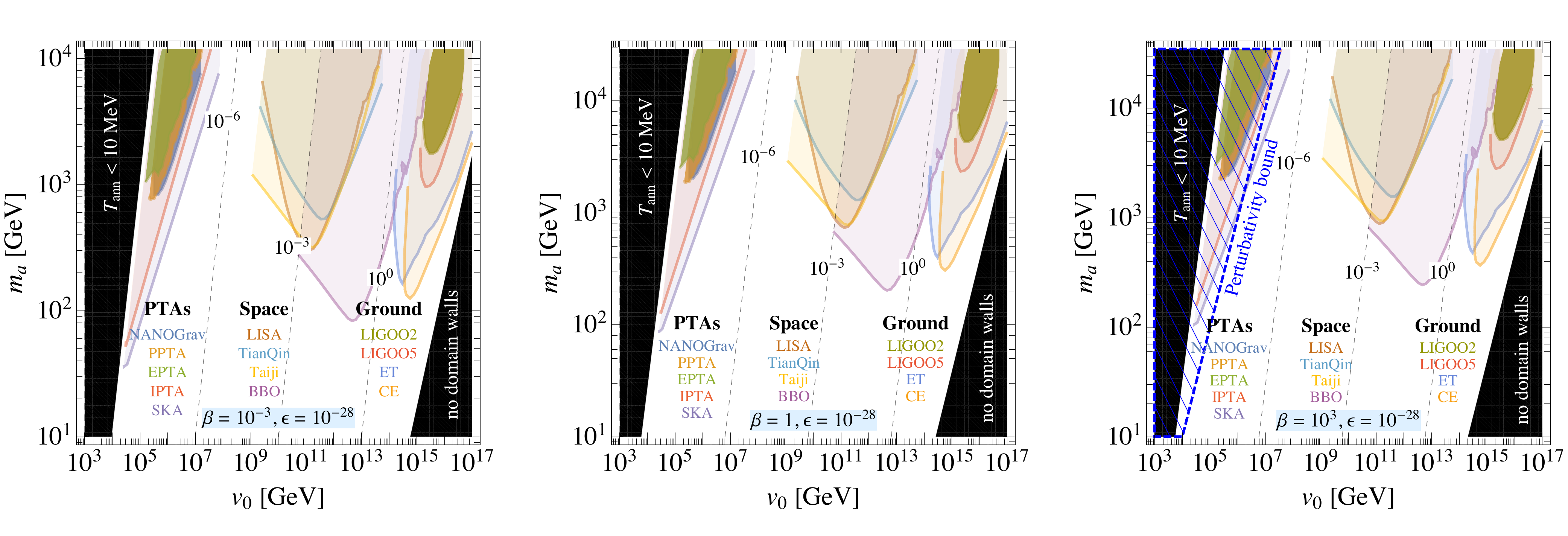}
\caption{The projections of reach of GW detectors in the $v_0$-$m_a$ plane, for fixed $\epsilon=10^{-28}$, and $\beta=10^{-3}$ (left), 1 (middle) and $10^3$ (right). The upper limits of $m_a$ during scanning are determined by the condition that domain walls should annihilate before they dominate the Universe. The parameter space with $T_{\rm ann}<10$ MeV or cannot form domain walls is covered by black shaded regions. The perturbativity bound is accounted with a dashed region. Peak frequencies of the GWs are plotted in black dashed contours.}\label{fig:gw}
\end{figure*}

In Fig.~\ref{fig:gw_f}, the sensitivity curves for various GW detectors are also shown, in which we have included the pulsar timing arrays (PTAs) NANOGrav~\cite{McLaughlin:2013ira,NANOGRAV:2018hou,Aggarwal:2018mgp,Brazier:2019mmu}, PPTA~\cite{Manchester:2012za,Shannon:2015ect}, EPTA~\cite{Kramer:2013kea,Lentati:2015qwp,Babak:2015lua}, IPTA~\cite{Hobbs:2009yy,Manchester:2013ndt,Verbiest:2016vem,Hazboun:2018wpv} and SKA~\cite{Carilli:2004nx,Janssen:2014dka,Weltman:2018zrl}, the space-based laser interferometers LISA~\cite{LISA:2017pwj}, TianQin~\cite{TianQin:2015yph,Hu:2017yoc,TianQin:2020hid}, Taiji~\cite{Hu:2017mde,Ruan:2018tsw} and BBO~\cite{Crowder:2005nr}, and the ground-based interferometers LIGO~\cite{LIGOScientific:2014qfs,LIGOScientific:2019vic}, CE~\cite{Reitze:2019iox} and ET~\cite{Punturo:2010zz,Hild:2010id,Sathyaprakash:2012jk}. The parameter space excluded by the accumulated data is covered by shaded  regions, while the projections of future detectors are shown in dashed lines. Given those sensitivity curves, we are able to perform a parameter scan on $v_0$, $m_a$, $\beta$ etc to see the reach of various GW detectors.

For simplicity, we assume one $V_{\rm bias}$ dominates the GW shape, and parametrise it as $\epsilon v_0^4$. We then fix $\epsilon=10^{-28}$, and scan over $v_0$ and $m_a$ for $\beta=10^{-3}$, 1 and $10^3$, and the results are given in Fig.~\ref{fig:gw}. The condition that the domain walls should annihilate before dominating the Universe is translated to
\begin{equation}
m_a<2.8\times10^4~{\rm GeV}\times\left(\frac{\epsilon}{10^{-28}}\right)^{1/2}\frac{1}{f(\beta)},
\end{equation}
in our parametrisation formalism, which determines the upper limit of $m_a$ in the scanning. On the other hand, the regions $T_{\rm ann}<10$ MeV are shaded in black as they are excluded by the Big Bang Nucleosynthesis. The regions that $V_{\rm bias}$ is too large such that large scale domain walls cannot form are also shaded in black. The peak frequency contours of the GW signals are plotted in straight black dashed lines. We can see that the GW astronomy is able to probe theories forming domain walls in a vast region of scale, from $v_0\sim{\rm TeV}$ up to as high as $10^{17}$ GeV.

\section{Conclusions \label{sec:V}}

A domain wall is a topological defect arising from spontaneous breaking of discrete symmetries. Spontaneous breaking of a global $Z_N$ symmetry in general leads to $N$ degenerate vacua. Domain walls, as barriers separating these vacua, can have different properties from those from $Z_2$ symmetry breaking.

We study on domain walls from $Z_3$ symmetry breaking, which have been rarely discussed in the literature.
From a general renormalisable $Z_3$-invariant potential, we obtain the relevant domain wall solutions numerically. We further derive semi-analytical solutions for the tension and thickness of the wall in terms of the vacuum expectation value, scalar mass, and a parameter $\beta$ which characterises the $Z_3$ property of the theory. Derivations from result in $Z_2$ domain walls are found, which are specified in terms of $\beta$. In the limit $\beta \to 0$, the $U(1)$ symmetry is recovered and no domain walls from after the SSB. In the small $\beta$ case, i.e., $\beta \ll 1$, our result is consistent with the domain walls studied in the axion model. In the large $\beta$ case, i.e., $\beta \gtrsim 1$, the tension  becomes $\beta$-insensitive.

We further discuss the gravitational radiation released from $Z_3$ domain walls with explicit breaking included. We show that as multiple degenerate vacua exist in the theory, an explicit breaking term leads to multiple biases between the vacua. Biases are classified into two types, depending on whether the wall separating the true vacuum and a false vacuum or two false vacua. Due to these biases, domain walls separating different vacua collapse at different time in the early Universe, and the process of domain wall collapsing is more complicated than the $Z_2$ case. As a consequence, GW spectrum from these walls is different from those in the $Z_2$ case. However, this effect is expected to be small as the longest-lived walls dominates the signal. We show that the future GW detectors, including PTAs and space- or ground-based interferometers have the potential to hit these domain walls in a very wide energy scale above the electroweak scale.

\acknowledgements
Y.W. thanks the U.S. Department of Energy for the financial support, under grant number DE-SC 0016013. K.P.X. is supported by the University of Nebraska-Lincoln.

\bibliographystyle{apsrev4-2}

\end{document}